# Two-dimensional universal conductance fluctuations and the electron-phonon interaction of topological surface states in $Bi_2Te_2Se$ nanoribbons

Zhaoguo Li[1], Taishi Chen[1], Haiyang Pan[1], Fengqi Song[*,1], Baigeng Wang[*,1], Junhao Han[1], Yuyuan Qin[1], Xuefeng Wang[2], Rong Zhang[2], Jianguo Wan[1], Dingyu Xing[1], Guanghou Wang[1]

[1]National Laboratory of Solid State Microstructures and Department of Physics, Nanjing University, Nanjing, 210093, P. R. China

[2]School of Electronic Science and Engineering, Nanjing University, Nanjing, 210093, P. R. China

Correspondence and requests for materials should be addressed to F. S. and B. W. (songfengqi@nju.edu.cn, bgwang@nju.edu.cn).

The universal conductance fluctuations (UCFs), one of the most important manifestations of mesoscopic electronic interference, have not yet been demonstrated for the two-dimensional surface state of topological insulators (TIs). Even if one delicately suppresses the bulk conductance by improving the quality of TI crystals, the fluctuation of the bulk conductance still keeps competitive and difficult to be separated from the desired UCFs of surface carriers. Here we report on the experimental evidence of the UCFs of the two-dimensional surface state in the bulk insulating $Bi_2Te_2Se$ nanoribbons. The solely-$B_\perp$-dependent UCF is achieved and its temperature dependence is investigated. The surface transport is further revealed by weak antilocalizations. Such survived UCFs of the topological surface states result from the limited dephasing length of the bulk carriers in ternary crystals. The electron-phonon interaction is addressed as a secondary source of the surface state dephasing based on the temperature-dependent scaling behavior.

In the new class of quantum condensed matter, namely three-dimensional (3D) topological insulators (TIs), the anomalous topological transition casts with a gapless and low-dissipation state on the solid surface[1,2]. The Dirac fermions within the surface states (SSs) are protected by the time-reversal symmetry[3]. Their spin helicity forbids backscattering of the Dirac carriers[4]. Robust coherence has been demonstrated in spite of the strong spin-orbit couplings in current 3D TI samples[5-9]. Such unique transport sheds light on the emergence of new mesoscopic physics[10]. The Aharonov-Bohm[11,12], Atshuler-Aronov-Spivak[5,13] and Aronov-Casher effect[14] as well as two-dimensional (2D) weak antilocalizations (WAL)[6-9] of the SSs have been demonstrated in TIs of $Bi_2Se_3$[15-18], $Bi_2Te_3$[15,19,20], $Bi_2Te_2Se$[21-23] and $Ag_2Te$[24], etc. These transport investigations provide the solid evidence on the conducting SSs and pave the way to the TI-based microdevices[25].

The universal conductance fluctuation (UCF)[26-28] is of intense interest in the TIs due to the expectation of the enhanced SS contributions as well as new fractional quantized amplitudes[29]. A main problem with TIs is the contribution from the bulk carriers[3,4,10], which is dominant over the SS transport even in the insulating TI samples due to their great numbers of electron states[20,22]. The UCF is featured by a reproducible aperiodic structure in the magnetoconductance (MC) curves of small weakly-disordered specimens. Such MC fluctuations are sensitive to the variations of disorder configurations and thus a "fingerprint" of the mesoscopic samples is visualized. One of the unique characteristics of the UCF is the nearly universal zero-temperature amplitudes in the order of $e^2/h$. For each individual transport channel, it keeps independent of the sample size and degree of disorders within the dephasing length $L_\varphi$[26,28]. In the UCF of TIs, both surface and bulk states are expected to have the comparable UCF amplitudes. Therefore, such unique UCF properties may result in a dominant SS's contribution when careful optimizations of the samples' transport environments are made[26]. This leads to great interests in the UCF measurements of TIs. The conductance fluctuations (CFs) from over 6 to 0.01 $e^2/h$ have been reported in $Bi_2Se_3$[30-32]. The CFs were even observed in well-refined macroscopic crystals[30]. One often determines the 2D character of the transport by field-tilting measurement due to the fact that the parallel field has no contribution to the phase[7]. However, such efforts on searching for the 2D UCFs remain challenging. In this report, we give the experimental evidence of the UCF effect in the bulk insulating $Bi_2Te_2Se$ nanoribbons. The 2D nature of the UCF is revealed by the field-tilting MC measurements.

**Results**

**Sample preparation and characterization.** The $Bi_2Te_2Se$ TI nanoribbons are prepared on a silica-capped silicon wafer by a mechanical exfoliation method. The four-probe electrodes were then applied onto the nanoribbons by a standard lift-off procedure. Fig. 1(a) shows the typical configuration during the transport measurement. The height ($H$), width ($W$) and length ($L$) of the present sample are 62 nm, 1.2 μm and 1.5 μm respectively, where L is defined as the distance between the voltage probes in the four-probe configurations. Fig 1(b) shows the temperature dependence of its resistance, indicating the bulk insulating of the sample. The energy gap Δ is 4.9 meV determined by adopting the Arrhenius equation[22]. The MC curves of the nanoribbons are measured from -9 to 9T while rotating the samples and varying the temperatures.

**Identifying the UCF features in the MC measurement.** The UCF patterns can be identified while the magnetic field (B) is perpendicular to the sample (θ=0°) as shown in Fig. 1(c). The conductance displays the aperiodic features and strongly temperature-dependent fluctuations. These features originate from the UCF effect as supported by the following characteristics. Firstly, the "noisy" (or aperiodic) CF patterns can be observed repeatedly. It shows the similar features in the different MC curves measured at the different temperatures because some specific "fingerprints" of the samples can be seen. Secondly, The root mean square of the CFs $\delta G_{rms} = \sqrt{\langle [\delta G(B) - \langle \delta G(B) \rangle]^2 \rangle}$, where $\langle \cdots \rangle$ expresses the ensemble average, also decays from nearly 0.008 to 0.002 $e^2/h$ when the temperature increases from 2 to 30 K, implying its quantum mechanical nature. It is well known that the UCF amplitudes undergoes the average reduction when the sample dimensions are longer than $L_\varphi$[26,28]. Quantitatively, following the calculations in Ref.[28,33], the 2D UCF theory predicts $\delta G_{rms} \sim 0.86(e^2/h)/(2N)^{1/2}$ at T=0K which agrees to the experimental value in orders, where $N = L \times W/L_\varphi^2$ represents the number of independent phase-coherent regions on the TI's surface. Finally, the temperature-dependent δGrms is investigated. According to the UCF theory for 2D system[26,28], we have $\delta G_{rms} \propto (\ln T/T)^{1/2}$, which also agrees to our experimental data, as shown in Fig 1(d). All the above experimental observations unambiguously support the UCFs in $Bi_2Te_2Se$ TI nanoribbons and suggest the 2D nature of the carriers.

**Experimental evidence of the 2D origin of the UCF.** We investigate the 2D origin of the UCF by the magnetic field-tilting MC measurements. As schematically shown in Fig 2(a), the phase shifts along some enclosed paths determine the UCF patterns upon the application of an external field[26,28,33]. In an ideal 2D electronic system, such phase shifts solely depend on the normal components ($B_\perp = B\cos\theta$) of the magnetic field. The UCF patterns accordingly evolve as a function of $B_\perp$. Therefore, the solely-$B_\perp$-dependent UCF patterns can manifest the UCF of a 2D electron system, essentially the novel SS for a TI sample[34]. It is indeed observed as shown in Fig. 2(b), where the angular-dependent UCFs are clearly seen. We can find three small peaks (p1, p2 and p3) in the $\delta G$-$B$ curves shift towards the high-$B$ direction and their widths are monotonically broaden with increasing $\theta$, as guided by the circle-marked lines. The locations of three peaks are plotted against $\theta$ in Fig. 2(c). The solid curves are the least square fittings, which ideally display the characteristic of $1/\cos\theta$ dependence. The 2D nature of the UCF pattern is further confirmed by the angle-dependent $\delta G_{rms}$, as shown in Fig. 2(d). Generally, $\delta G_{rms}$ is expected to be unchanged while varying $\theta$ due to the isotropic $L_\varphi$ in a 3D system. We can find $\delta G_{rms}$ maintains comparable while $\theta$ is below 40°. However, when $\theta$ exceeds 45°, $\delta G_{rms}$ drops abruptly, which can be explained by the contribution of a 2D conducting states. In a TI system, the UCF contributions from the electrons of the bulk states gradually predominate while θ is increasing[17,20]. The present anisotropic $\delta G_{rms}$ rules out the 3D origin. We confirmed the $\theta$-variable MC measurements in other samples, proving the generality of the above evidence. Both the solely-$B_\perp$-dependent UCF patterns and anistropic UCF amplitudes further conclusively demonstrate the 2D UCF of the topological SSs in $Bi_2Te_2Se$.

**The WAL features dominated by surface carriers.** The SS transport in our samples is further examined by the solely-$B_\perp$-dependent WAL effect. Fig. 3(a) shows the angular-dependent magnetoresistance (MR) at 2K. It can be seen that the WAL-characteristic MR dips gradually disappear with increasing $\theta$. At $\theta = 90°$, where $B$ is along the current direction, the MR dip completely disappears and the MR curve shows a parabolic $B$ dependence. Such semiclassical $B^2$ dependence comes from the Lorentz deflection of the bulk carriers. According to the previous analysis[7,8], we plot $\Delta G$ against $B_\perp$, where all the MC curves at various angles coincide with each other [Fig. 3(b)], strongly confirming the transport of the 2D SSs[7]. The temperature dependence of the WAL features at $\theta = 0°$ has also been measured as shown in Fig. 3(c). According to the 2D localization theory, in the limit of strong spin-orbit interaction and low mobility, the Hikami-Larkin-Nagaoka equation is reduced to[6,7,35]:

$$\Delta G(B) \approx -\alpha \frac{e^2}{2\pi^2 \hbar} \left[ \psi\left(\frac{1}{2} + \frac{B_\varphi}{B}\right) - \ln\left(\frac{B_\varphi}{B}\right) \right] \qquad (1)$$

where $\alpha=1/2$ stands for the transport through only one TI surface channel[36], $\psi(x)$ is the digamma function, and $B_\varphi = \frac{\hbar}{4eL_\varphi^2}$ is a $L_\varphi$-related characteristic field. By applying Eq. (1) to the experimental curves, the fitting parameters of $\alpha$ and $B_\varphi$ are obtained. Fig. 3(d) shows the fitting results of $\alpha$, whose value around 0.5 can be seen. This indicates that only a single surface channel participates in the transport of our TI samples.

**Discussions**

$L_\varphi$ characterizes the mesoscopic electron interference and its temperature-dependence may reveal the phase relaxation mechanism of the SSs. The temperature-dependent scaling of $L_\varphi$, extracted from either WAL or UCF data, both present a similar trend as shown in Fig 4. In the previous framework of WAL and UCF[26-28,37], the electron-electron (e-e) scattering and electron-phonon (e-ph) scattering are considered as two main sources of the electrons' phase relaxation. The e-e interaction theory predicted a power law of $L_\varphi \propto T^{p'/2}$, where $p'= 1$ for the 2D system[27,37], as employed in the previous studies of graphene[38] and TIs[8,32]. The dashed line in Fig. 4 is the $T^{-1/2}$ power-law curve following the 2D e-e interaction, which clearly shows the failure of the fitting. Even if we optimize the parameter $p'$ it still does not work. This indicates that our data can not be explained by the e-e interaction dominated dephasing mechanism.

Here we propose an e-ph interaction dependent scaling behavior for the dephasing length $L_\varphi$ of the SS electrons in the TIs. It is well-known that the e-e scattering used to be dominant in the 2D system, while the e-ph scattering predominates in the 3D system[37]. However, as compared to graphene, an ideal 2D electronic system, a huge phonon "sea" attaches to the SS electrons in a TI. The e-ph interaction is supposed to emerge in the present study[39], which contributes an additional term in the formula as shown in the following[37]:

$$\frac{1}{L_\varphi^2(T)} = \frac{1}{L_{\varphi 0}^2} + A'_{ee} T^{p'} + A'_{ep} T^{p} \qquad (2)$$

where $L_{\varphi 0}$ represents the zero-temperature dephasing length, $A'_{ee}T^{p'}$ and $A'_{ep}T^{p}$

represents the contribution from the e-e and e-ph interaction respectively. We have known $p'=1$ according to the above analysis. In the current samples of the Bi$_2$Te$_2$Se nanoribbons, the electron-transverse phonon (e-t.ph) interactions should be dominant over the electron-longitidunal phonon (e-l.ph) interactions because of the larger sound velocity of longitidunal phonons than that of transverse phonon in the material[40,41]. According to e-ph interaction theory[41,42], the e-ph scattering length $L^{-2}_{e\text{-}ph} \propto T^p$. The e-t.ph scattering leads to the value of $p=2$ while the e-l.ph interaction yields $p=3$[41]. Therefore, p is close to 2 in our samples. As shown in Fig. 4, the Eq. (2) with $p'=1$ and $p = 2$ results in a perfect fitting with the parameter $L_{\varphi 0} = 114$ nm. Such a fitting formula is applicable for all of our samples. This highlights the e-t.ph interaction in the electron relaxation of the present TI samples.

The above study of the transport environment well explains the present observations of the solely-$B_\perp$-dependent 2D UCF. In TIs, the UCF patterns are composed of contributions from the SS and bulk carriers, both of which undergo the average reduction of their amplitudes while the sample size exceeds $L_\varphi$. A poor-man approach to obtain the SS-dominated UCF is to suppress the amplitudes of the UCF from bulk carriers, as has been implemented in this work. One may know δG$_{rms}$ is dominantly contributed by the bulk electrons while measuring the MC curve at some high angles near $\theta=90°$. Such UCF amplitudes from the bulk electrons should be isotropic for a 3D interference system. In our samples, $\delta G_{rms}$ ($\theta = 90°$) equals 0.003 e$^2$/h, which is much lower than the amplitudes of SS UCF. This confirms the suppression of the bulk UCF firstly. Secondly, $L_\varphi^{3D} \sim 12$ nm can be estimated with a comparable error by applying the 3D UCF theory[28]. All of our samples exhibit the similar bulk dephasing lengths. Due to the fact that $L_\varphi^{3D} < H, L, W$, we believe the 3D interference transport of the bulk electrons in our samples. Accordingly, the 2D UCF discussed above reasonably originates from the topological SSs. Thirdly, we see the transport quality of the samples. Generally, the TI samples have been delicately prepared to keep the perfection of the crystals in order to maintain the topological anomaly and the well-defined SS transport[3], which leads to some high values of bulk mobilities, especially for the binary TIs. This accordingly expects a satisfactory $L_\varphi$ and a UCF amplitude for the bulk electrons competitive to those of the SS electrons. However in the ternary TIs, high defect ratios and low mobilities have been demonstrated[23] due to plenty of antisite defects between Te and Se and related vacancies. A much lower $L_\varphi$, i.e. about 10nm, for the bulk carriers is then reasonable, while a relatively good value, i.e. about 120nm, is obtained for the SS since the SS carriers may immune to some scatterings due to their helical transport[3,4]. Therefore, we finally argue that the "deteriorated" electronic environment in the ternary TIs suppresses the bulk UCF and helps the survival of the 2D UCFs in the present study.

In summary, we have successfully demonstrated the UCF of the 2D topological SSs by measuring the solely-$B\perp$-dependent UCF patterns. It is observed in the TI samples with a short L$\varphi$ of the bulk carriers, which suppress the UCF's amplitudes from the bulk carriers and helps the survival of the UCF of the 2D topological SS. The e-ph scattering is suggested as the other dephasing source for the SS of TI samples. The present work may pave the TI-based spintronic devices and quantum information materials and their room-temperature applications.

**Methods**

**Crystal growth and characterization.** The well-refined $Bi_2Te_2Se$ crystals were grown by melting the high purity powders (99.999%) of Bi, Te, Se with a molar ratio of 2:2:1 at 850℃ in evacuated quartz tubes for 3 days. It was followed by cooling slowly to 550℃ for 8 days and then an annealing for 5 days before rapidly cooling to room temperature. The ordering of the chalcogen layers in the $Bi_2Te_2Se$ precursor crystals was confirmed by the x-ray powder-diffraction[22]. The $Bi_2Te_2Se$ nanoribbons were then exfoliated and transferred onto the wafers[5]. The four-probe electrodes were then applied onto the nanoribbons by a standard lift-off procedure.

**Transport measurements and data analysis.** The transport measurements were carried out in a Quantum Design Physical Property Measurement System-9T system and a homemade MR measurement system. The MR curves of the nanoribbons were measured from -9 up to 9T while rotating the samples from 0 to 90° and changing the temperatures from 2 to 300K. The UCF features are always mixed with the WAL dips. We first fit the MC curves according to the traditional WAL description. After subtracting the smooth background, we obtained the resultant aperiodic structures as the UCF features as discussed in the main text. The dephasing length can be extracted from both WAL and UCF effect. In the WAL description, the dephasing length $L_\varphi^{WAL}=[\hbar/(4eB_\varphi)]^{1/2}$ was obtained by fitting the characteristic field. In the UCF theory, the correlation function $F(\Delta B)=<\delta G(B)\delta G(B+\Delta B)>$ must be calculated with the result of a critical field $B_c$ satisfy $F(B_c)=F(0)/2$. Using $(L_\varphi^{UCF})^2 \times B_c \sim h/e$, we then reach the dephasing length from UCF data. The UCF features decayed faster than the WAL dips with the increase of the temperature and nearly vanished at 30K. Therefore, the $L_\varphi^{UCF}$ data presents of larger error and $L_\varphi^{WAL}$ will be more discussed.

**Acknowledgments:** We thank the National Key Projects for Basic Research of China (Grant numbers: 2010CB923400, 2011CB922103), the National Natural Science Foundation of China (Grant numbers: 11023002, 11134005, 60825402, 61176088, 11075076), the PAPD project and the Fundamental Research Funds for the Central Universities for financially supporting the work. Helpful assistance from Nanofabrication and Characterization Center at Physics College of Nanjing University, Yanfang Wei and Prof. Zhiqing Li at Tianjin University, Prof. J. J. Lin at Taiwan Chiao Tung University, Dr. Li Pi and Prof. Yuheng Zhang at High Magnetic Field Laboratory CAS are acknowledged.

**Figure Caption:**

**Figure 1**. The UCF and its temperature dependence. (a) the schematic diagram of the measurement configuration. (b) Temperature dependence of the resistance and resistivity of a $Bi_2Te_2Se$ nanoribbon. The left inset shows its AFM image with the scale bar of 4μm. The right inset shows the Arrhenius fitting of ρ(T) with the result of a 4.9 meV band gap. (c) Conductance fluctuations plotted against B at various temperatures (θ=0). The aperiodic R-B patterns appear repeatedly. (d) $δG_{rms}$ and its temperature dependence. The inset shows the data in a linear scale. The solid curve is fit by the traditional UCF theory.

**Figure 2**. The 2D UCFs demonstrated by the field-tilting measurement. (a) The schematic diagram showing the 2D UCF solely depends on the perpendicular component of the magnetic field ($B_⊥$). (b) The *B*-tilting δ*G*-*B* data of a $Bi_2Te_2Se$ nanoribbon measured at 2 K. The black, red and blue circle-marked lines respectively show the similar features, namely p1, p2 and p3, in all the δ*G*-*B* curves. (c) The positions of the UCF features plotted against *θ*. The black, red and blue data are from those of p1, p2 and p3, respectively. The solid curves are the 1/cos*θ* fitting. (d) *θ* dependent $δG_{rms}$. The dashed curve is for eye guiding. The UCF measured at *θ* = 90° is interpreted as the contribution from the bulk carriers in TIs.

**Figure 3**. The 2D WAL effect of the SSs in TIs. (a) Δ*R* = *R*(*B*) - *R*(0) as a function of *B* measured at various *θ* at 2 K. (b) Δ*G*($B_⊥$) curves after subtracting the bulk contribution. The data were measured at *T* = 2 K. (c) Δ*G*(*B*) as a function of *B* at various temperatures (*θ* = 0). The solid curves are the least-square fittings according

to the 2D WAL theory. (d) $\alpha$ as a function of temperature obtained from the fitting. Its value near 0.5 rules out the issue of electron interference between two surfaces.

**Figure 4**. Temperature dependence of $L_\varphi$ obtained from the WAL (square) and UCF data (circle). The big error of $L_\varphi$ at 30 K from UCF is due to nearly disappearance of the UCF. The solid curve shows the fitting of the square-dotted curve according to the scaling formula (2) with $p' = 1$ and $p = 2$. The dashed line is the e-e interaction fitting.

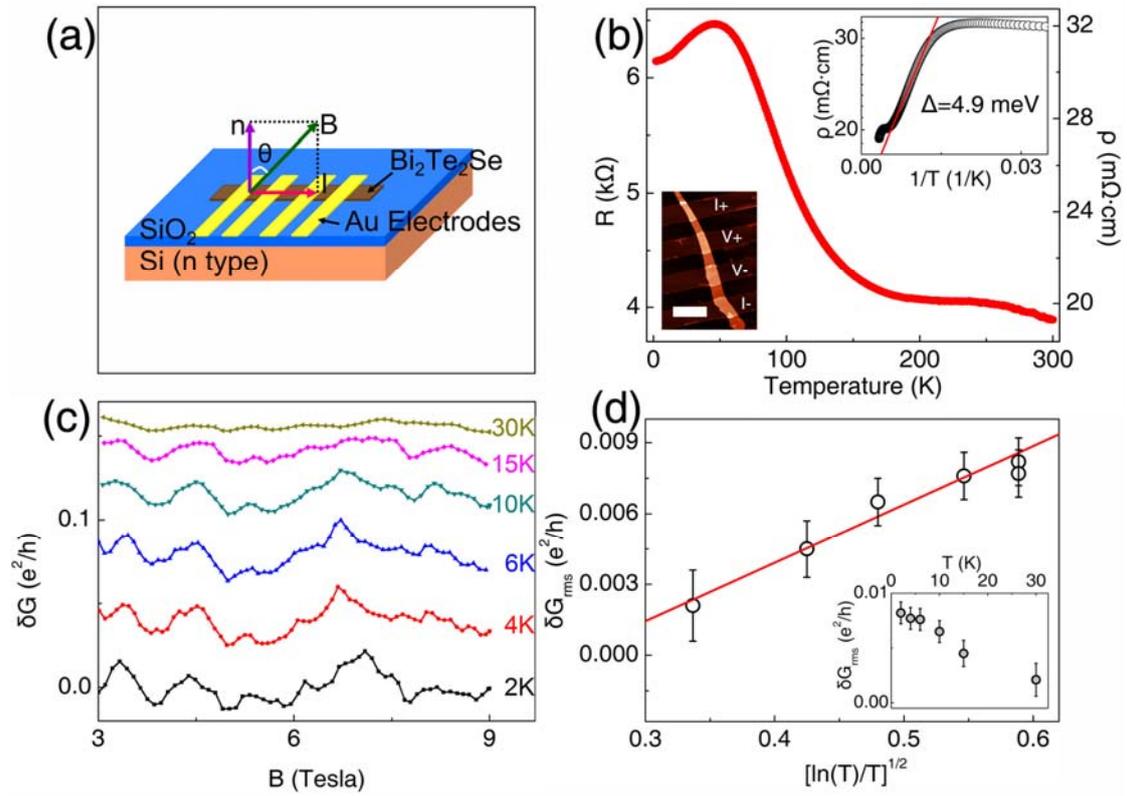

Figure 1 Li et al

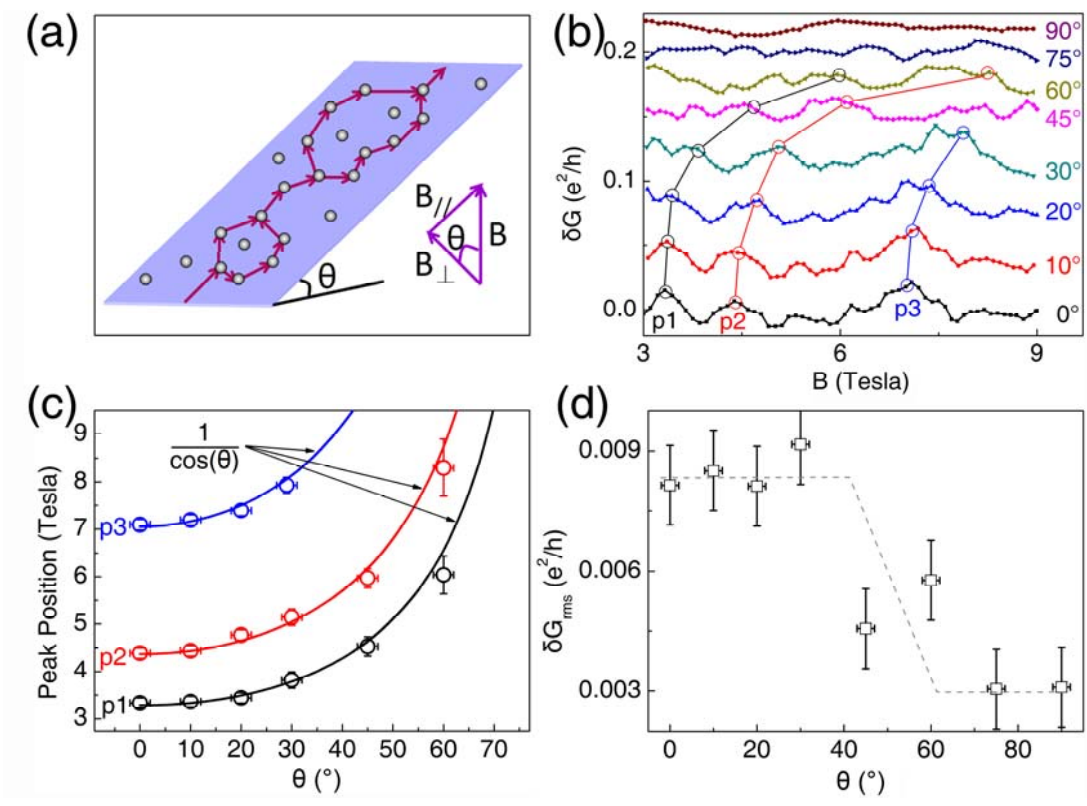

Figure 2 Li et al

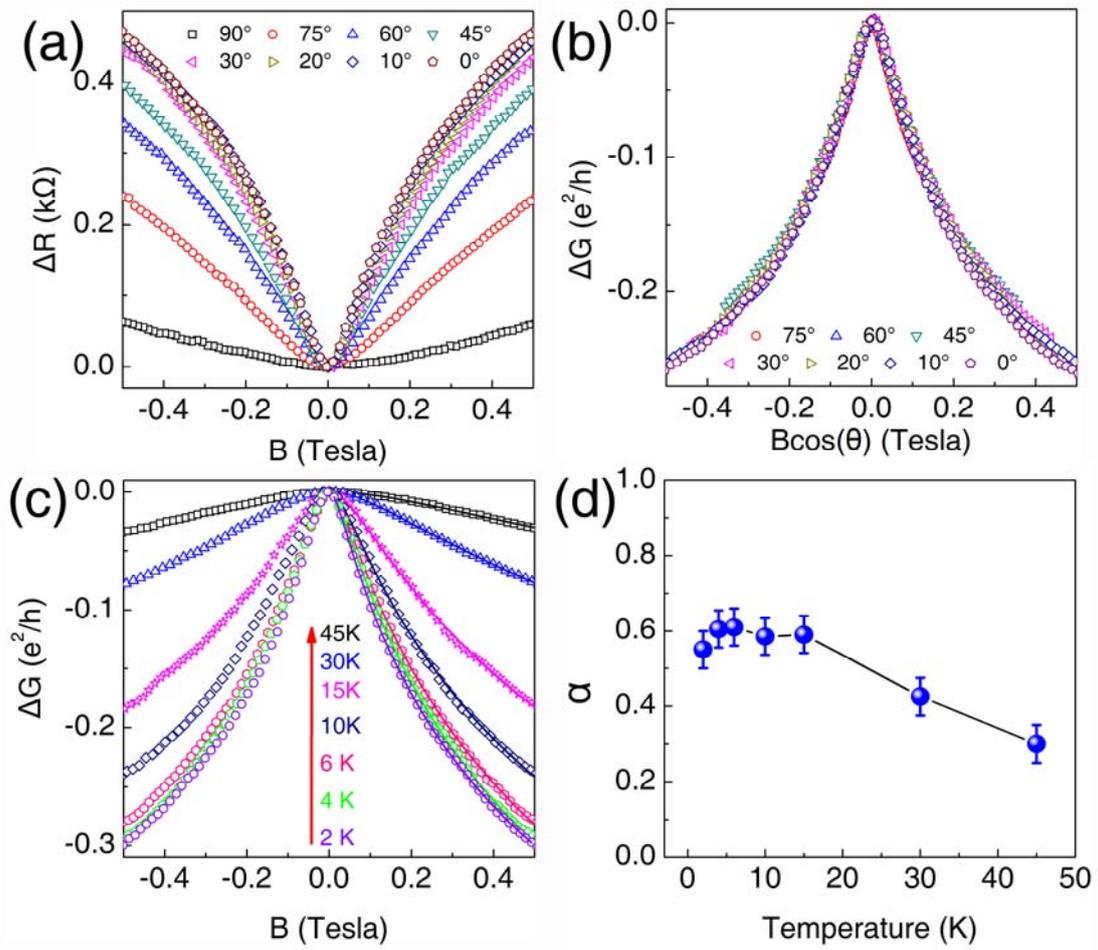

Figure 3 Li et al

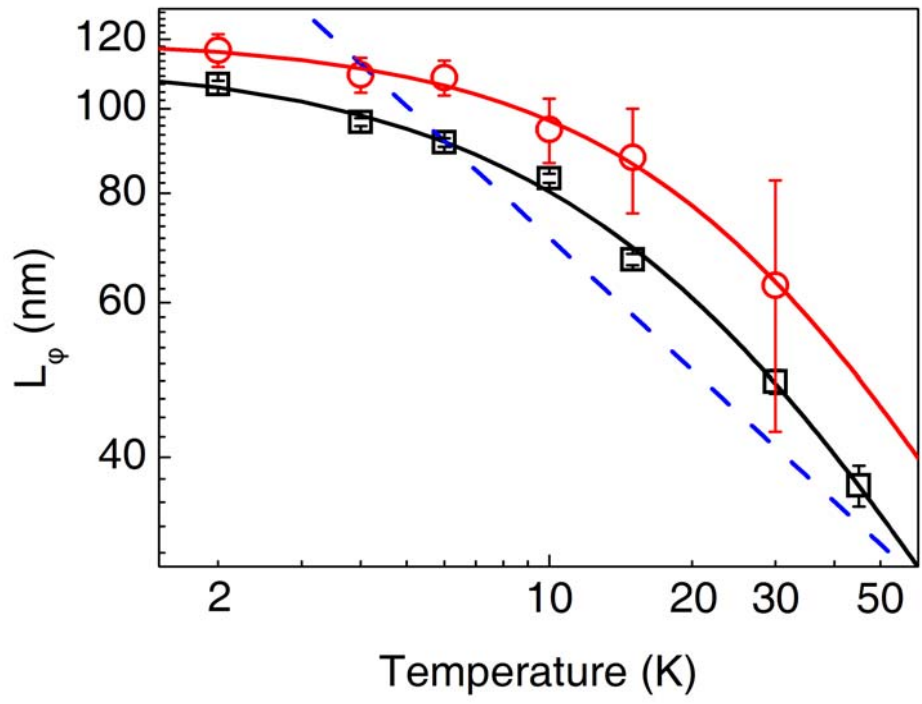

Figure 4 Li et al